\documentclass[screen,natbib,nonacm,sigconf]{acmart}

\usepackage[table]{xcolor} 
\usepackage{hhline}        
\usepackage{array}         
\usepackage{multirow}      
\usepackage{booktabs}      
\usepackage{colortbl}      
\usepackage{graphicx} 
\usepackage{physics}
\usepackage{amsfonts}
\usepackage{amsthm}
\usepackage{amsmath}
\usepackage{multirow}
\usepackage{bbm}
\usepackage{cancel}
\usepackage{enumitem}
\usepackage{soul}
\usepackage{slashed}
\usepackage{nicefrac}

\usepackage{orcidlink}
\usetikzlibrary{decorations.pathmorphing,arrows.meta}
\usepackage{physics}
\usepackage{amsthm}
\usepackage{mathtools}  
\usepackage{csquotes}

\newcommand{\B}{B}

\newcommand{\bl}{b}

\newcommand{\I}{\mathbb{I}}

\newcommand{\GR}[2][{}]{\ensuremath{\mathbb{R}^{{#1}}_{#2}}}
\newcommand{\GC}[2][{}]{\ensuremath{\mathbb{C}^{{#1}}_{#2}}}
\newcommand{\e}{\mathbf {e}_}            
\newcommand{\ebar}{\mathbf {\bar{e}}_}            

\newcommand{\Spin}[2][{}]{\ensuremath{\text{Spin}^{#1}({#2})}}

\newcommand{\Pin}[2][{}]{\ensuremath{\text{Pin}^{#1}({#2})}}
\newcommand{\SU}[2][{}]{\ensuremath{\text{SU}_{#1}({#2})}}
\newcommand{\UN}[2][{}]{\ensuremath{\text{U}_{#1}({#2})}}

\newcommand{\spin}[1]{\ensuremath{\mathfrak{spin}({#1})}}
\newcommand{\su}[2][{}]{\ensuremath{\mathfrak{su}_{#1}({#2})}}

\newcommand{\un}[2][{}]{\ensuremath{\mathfrak{u}_{#1}({#2})}}

\usepackage{hyperref}
\usepackage{cleveref}

\definecolor{Xcolor}{HTML}{fff0eb}
\definecolor{Scolor}{HTML}{fff0eb}  
\definecolor{Stildecolor}{HTML}{fff0eb}
\definecolor{nuRcolor}{HTML}{edf4ff}  
\definecolor{epsilonLcolor}{HTML}{fee5f0}  
\definecolor{nuLcolor}{HTML}{feeef2}  
\definecolor{epsilonRcolor}{HTML}{e1ecff}  
\definecolor{headercolor}{HTML}{f7feeb}

\newcommand{\rrho}{\boldsymbol{\rho}}

\begin{document}

\title{Lepton Triptych I: Geometric Foundations of Electroweak Symmetry in the Real Clifford Algebra \texorpdfstring{$\text{Cl}_4(\mathbb{R})$}{Cl₄(ℝ)}}


\author[Roelfs]{Martin Roelfs}
\authornote{Both authors contributed equally to the paper.}
\orcid{0000-0002-8646-7693}
\affiliation{%
  \department{Department of Mathematics}
  \institution{University of Antwerp}
  \city{Antwerp}
  \country{Belgium}
}
\email{martin.roelfs@uantwerpen.be}
\author[Eelbode]{David Eelbode}
\authornotemark[1]
\orcid{0000-0002-2919-262X}
\affiliation{%
  \department{Department of Mathematics}
  \institution{University of Antwerp}
  \city{Antwerp}
  \country{Belgium}
}
\email{david.eelbode@uantwerpen.be}

\date{\today}

\begin{abstract}
This paper investigates how the spinor space of the electroweak gauge group $\text{SU}_{I}(2) \times \text{U}_{Y}(1)$ can be derived using recent geometric techniques within the real Clifford Algebra $\mathbb{R}_4 = \text{Cl}_4(\mathbb{R})$.
Central to this approach is a novel procedure for constructing the spinor space of $\mathbb{R}_4$ directly, without complexification or matrix representation.
In fact, the defining projector of the spinor space 
corresponds to the zero matrix in the defining representation of $\text{SU}(2)$, and hence this construction has no $2 \times 2$ complex matrix counterpart.

We subsequently show that the spinor space of $\mathbb{R}_4$ naturally accommodates irreducible representations (irreps) for a single generation of chiral Standard Model leptons, including a sterile right-chiral neutrino.
Left- and right-chiral particles arise due to the grade-parity of the irreps, explaining geometrically why weak isospin acts exclusively on left-chiral states.
This manuscript shows how the real Clifford algebra $\mathbb{R}_4$ is the smallest algebra to house irreps for both the $\text{SU}_{I}(2) \times \text{U}_{Y}(1)$ gauge bosons and the Standard Model leptons.
Simultaneously the geometric approach presents a way to compute interactions that does not depend on irreps, but which instead allows computations on the higher level of bosons and leptons.

The emergence of the correct interactions directly from first principles highlights the promise of this framework not only for the geometric foundations of Electroweak Theory, but also for the Standard Model and Grand Unified Theories more broadly.
This paper is the first panel of the Lepton Triptych, which will ultimately present the full Yang-Mills theory of the electroweak model based on these principles.
\end{abstract}


\maketitle

\section{Introduction}

This work is the first panel in our Lepton triptych, where we develop Electroweak Theory from the ground up by following geometric principles.

In this first panel we will investigate how the defining gauge group $\UN{2} = \SU{2} \times \UN{1}$ of Electroweak Theory can be formulated from purely group theoretical principles in the (real) Euclidean space \GR[4]{}, and will find that this naturally leads to spinor representations which behave like lepton states.
The group \UN{2} is a subgroup of \Spin{4} \cite{LGasSG}, which is
itself the even subgroup of \Pin{4}; the group of all reflections in volumes through the origin in \GR[4]{}. 
In order to algebraically model the composition of reflections as dictated by the Cartan-Dieudonné theorem, we need an algebraic structure that captures the behaviour of reflections.
This is exactly what a Geometric Algebra\footnote{Geometric Algebras are alternatively known as Clifford Algebras, see \cite{doran} for a history of the usage of both terms.} (GA) provides, since reflections are anti-commuting involutions, just like the basis vectors of a GA \cite{GSG}.
The authors of \cite{GSG} have therefore previously said
that an algebra $\GR{p,q,r} \coloneqq \mathcal{G}(\GR[p,q,r]{}) = \text{Cl}_{p,q,r}(\mathbb{R})$ directly \emph{realizes} \Pin{p,q,r}, and by considering all elements of \GR{p,q,r} as compositions of reflections we can directly find various \emph{representations} for \Pin{p,q,r} to act on, corresponding to points, lines, planes, rotations, translations, screws, etc. \cite[Section 10]{GSG}.

Algebraic spinors however, remained elusive as they typically require \emph{complexification} when $d \geq 4$, in this particular case from $\GR{4}$ to $\GC{4}$, which placed them somewhat outside of the geometric interpretation developed in \cite{GSG,DKER2024}.
However, in this work we present a novel construction of abstract spinors in $d = 4$ that does not require complexification and the construction of a maximally isotropic subspace, but that is nonetheless isomorphic to the abstract spinors obtained by authors such as Polchinski \cite{Polchinski:1998rr} and Jürgen Jost \cite{alma9912723332602411}. As such, our approach bridges these more traditional methods using complex algebraic spinors and earlier attempts \cite{McClellan2017,Hestenes1982,GA4Ph} to develop particle physics using a real GA, by developing \emph{real} algebraic spinors in a real geometric setting.
In so doing, we find that \GR{4} naturally houses \emph{chiral} representations under \UN{2} that we recognize as the chiral leptons of the Standard Model.
Interestingly, our geometric model predicts a right-handed neutrino that couples to neither \SU[I]{2} nor \UN[Y]{1}, but that can nonetheless participate in the Higgs mechanism to provide mass to the neutrino.

In the second panel we will apply the same geometric principles to Minkowski space \GR[1,3]{} in order to realise the spinor space in \GR{1,3}, which has several challenges that are not faced in the purely Euclidean algebra \GR{4}, and therefore requires a separate paper to be dealt with correctly.
The resolution of these challenges naturally leads to answers to questions about chirality and Weyl spinors, charge conjugation and Majorana spinors, the geometric foundations of gauge invariance, and much more.

The third and final panel of the triptych will combine the first two into the Yang-Mills theory known as Electroweak Theory or Glashow-Weinberg-Salam theory. 
Here we shall reap the rewards of our efforts, because the geometric formulation of Electroweak Theory will naturally explain why the Standard Model leptons couple to hypercharge \UN[Y]{1} and weak isospin \SU[I]{2} with the coupling strength that they do, why left and right chiral particles behave differently under \SU[I]{2}, why the right-handed neutrino does not appear in the Standard Model Lagrangian but might still exist, why the Higgs mechanism takes the form that it does, and more.

This work is also the first in an even larger effort to develop Geometric Particle Physics. 
Future papers will include an extension of the principles of the current paper to the gauge group \SU{3}, 
as well as a more general classification of spinor spaces in geometric algebras $\GR{p,q}$ without need for complexification.

The ideas in this paper are most naturally expressed using the language of Geometric Algeba (GA). In fact, one of the central findings of this paper is that the spinor spaces appear naturally from group theoretic principles in a way that literally cannot be captured by the defining matrix representation.
We assume that the reader is familiar with the basics of geometric algebra; some good introductory texts for physicists are \cite{STA,GA4Ph}.

This article is organized a follows. \Cref{sec:un2} introduces the unitary group \UN{2} as a subgroup of \Spin{4} in \GR{4}. \Cref{sec:spineors} explores the spinor space of \GR{4} and clarifies the distinction between left and right chiral spinors. \Cref{sec:spinorproducts} explores the products between spinors. \Cref{sec:smfermions} then summarizes the representations of leptons and gauge bosons in this model. Lastly, \cref{sec:discussion} gives an elaborate discussion of the main results of this paper and outlines future research.

\section{The Unitary Group \texorpdfstring{$\UN{2} = \SU{2} \times \UN{1}$}{U(2) = SU(2) ⨯ U(1)}}\label{sec:un2}

A group element $R \in \Spin{4}$ is the composition of at most four reflections in volumes through the origin, and hence of the form $R = v_4 v_3 v_2 v_1$, where $v_j \in \GR[4]{}$ is a reflection satisfying $v_j^2 = 1$. Equivalently, the group element $R$ is the exponential of a Lie algebra element $B \in \spin{4}$, where $B \in \GR[(2)]{4}$ is a bivector.
In general the generating bivector $B$ can be decomposed into simple ($\bl_j^2 \in \mathbb{R}$) commuting bivectors $\bl_j$ as $B = \tfrac{1}{2}(\theta_1 \bl_1 + \theta_2 \bl_2)$, 
where $\theta_j \in \mathbb{R}$ are the angles of rotation around the (invariant) axes $\bl_j$. In the particular case that the angles satisfy $\theta_1 = \pm \theta_2$, the rotation is called an \emph{isoclinic rotation}, and the decomposition of $B$ into $\bl_1$ and $\bl_2$ is no longer unique, and the stabilizer group of $B$ becomes \UN{2} \cite{LGasSG}.

To explore this in more detail, consider an isoclinic rotation $U = \exp(\theta \rrho_0)$ in at least 4D, generated by the bivector
    \begin{equation}
        \rrho_0 = \tfrac{1}{2} (b_1 + b_2),
    \end{equation}
with $b_1^2 = b_2^2 = -1$ and $b_1 b_2 = b_2 b_1$.
The commuting $2$-blades $b_j$ can then be factored into orthogonal vectors $\e{j}$ and $\ebar{j}$:
\[ b_j = \e{j} \ebar{j} = \e{j} \wedge \ebar{j}\ . \]
These vectors are the generators of a $4$ dimensional geometric algebra \GR{4}.
Since each $\bl_j$ squares to $-1$ and thus behaves like an imaginary unit, the bivector $\rrho_0$ is somewhat reminiscent of the diagonal matrix $i\mathbbm{1}$, and defines a complex structure on the vector space \GR[4]{} \cite{LGasSG}.
This bivector, which forms the backbone of this paper, will be referred to as the \emph{spine} throughout this paper.

The stabilizer group of the spine $\rrho_0$ under conjugation is the unitary group $\UN{2}$ \cite{LGasSG}, and therefore the elements of the corresponding Lie algebra $\un{2}$ must be bivectors that commute with $\rrho_0$. Apart from $\rrho_0$ itself, there are an additional three such bivectors:
    \begin{align}\label{eq:paulibivectors}
        \rrho_1 = \tfrac{1}{2} (\e{1}\e{2} + \ebar{1}\ebar{2}), \,
        \rrho_2 = \tfrac{1}{2} (\e{1}\ebar{2} - \ebar{1}\e{2}), \,
        \rrho_3 = \tfrac{1}{2} (\bl_1 - \bl_2).
    \end{align}
These additional three bivectors are the generators of $\SU{2}$ and correspond to the Pauli matrices $i \sigma_a$ \cite[Ch. 6]{roelfs_thesis}, and henceforth will be referred to as the Pauli bivectors.

While the scalar $1$ is the trivial identity element in \GR{4}, there is another identity element that is more deeply associated with $\SU{2}$ in specific.
To find it, consider the squares of the Pauli bivectors:
\begin{alignat*}{3}
	\rrho_a^2 &= - \tfrac{1}{2}\left(1 + \bl_1\bl_2 \right) 
    \quad &\leftrightarrow \quad (i \sigma_a)^2 = - \smqty(1 & 0 \\ 0 & 1 )
\end{alignat*}
where we have also shown the {matrix equivalent} if the standard Pauli matrices $i \sigma_a$ are used instead.
The element
\[ \I \coloneqq \tfrac{1}{2}(1 + \bl_1 \bl_2) \]
is a non-trivial identity element for \SU{2}. Indeed, it is different from the identity $1$ of the algebra \GR{4} but because it satisfies $\I^2 = \I$ and $\I \rrho_a = \rrho_a \I = \rrho_a$ for $a = 1, \ldots, 3$, the element $\I$ indeed behaves like an identity element for the group $\SU{2}$. 
In fact, as we have seen above it is $\I$ that maps to the identity matrix $\mathbbm{1} \in \mathbb{R}^{2 \times 2}$ when Pauli matrices are used by letting $\rrho_a \to i \sigma_a$, and \emph{not} the multiplicative identity $1$ of the group \Spin{4} in \GR{4}.
But if $\I$ and $1$ are both idempotents that behave like an identity for \SU{2}, then what about 
\[ \Box \coloneqq 1 - \I \; ?  \]
(Pronounced as ``box''.)
After all, this is also an idempotent satisfying $\Box^2 = \Box$, but evidently satisfies $\rrho_a \Box = \Box \rrho_a = 0$ for $a \in \{1,2,3\}$ and $\I \Box = \Box \I = 0$.
Another way in which both $\I$ and $\Box$ naturally appear is by exponentiating a Lie algebra element $\rrho = \sum_{a=1}^3 \theta_a \rrho_a \in \su{2}$ to form an $\SU{2}$ element
    \begin{equation*}
        e^{\rrho} = \Box + \I \cos(\theta) + \hat{\rrho} \sin(\theta) \quad \leftrightarrow \quad \mathbbm{1} \cos(\theta) + i \hat{\sigma} \sin(\theta) \ ,
    \end{equation*}
where $\theta = \sqrt{- 2 \rrho \cdot \rrho}$ and $\hat \rrho = \rrho / \theta$ such that $\hat{\rrho}^2 = - \I$, see \cite[6.4.2]{roelfs_thesis}.
This again makes it clear that $\I$ should be identified with $\mathbbm{1}$, but for any $\rrho \in \su{2}$ there is \underline{always} a constant term $\Box$ \emph{for any} \SU{2} element, which does not appear in the matrix representation.
All of these properties taken together show that as a matrix $\Box$ behaves like the \underline{null matrix}, but here we clearly have $\Box \neq 0$!
So what kind of sorcery is this?

\subsection{The Forgetful Idempotent}\label{sec:forgetful}

The $\bl_j$ are distinguishable elements which behave like commuting imaginary units. So as alluded to earlier the spine $\rrho_0$ is not quite the same as the matrix $i \mathbbm{1}$, precisely because of this distinguishability. This thus raises an interesting question: what if we were somehow unable to distinguish these different imaginary units $\bl_j$, and instead they all appear to us like the same imaginary unit $i$?

It is precisely this feat of selective amnesia that can be achieved by $\Box = 1 - \I$, which can be alternatively defined as
    \begin{align}
        \Box &\coloneqq{} (1 - \I) = \tfrac{1}{2} (1 - b_1 b_2) \ .
    \end{align}
It is straightforward to check that $\Box$ cannot distinguish between the $-b_1 b_2$ and $1$, in other words we have
    \begin{equation}
        \Box = - b_1 b_2 \Box \ .
    \end{equation}
Furthermore, because $\Box$ can always regurgitate $-b_1 b_2$ at any time, we find that e.g. $b_1 \Box = - b_1 b_1 b_2 \Box = b_2 \Box$. This leads to the equality
    \begin{equation}
        b_1 \Box = b_2 \Box .
    \end{equation}
The forgetful idempotent $\Box$ therefore has the desired property that we can no longer distinguish different commuting imaginary units and instead, they all appear the same to us.
The spine $\rrho_0$ is the complex structure because it satisfies
    \[ \rrho_0^2 \Box = - \Box \ .  \]
Following similar logic, all elements of the $2^4$ dimensional geometric algebra \GR{4} can be shown to fall into groups of two that can no longer be distinguished.
Explicitly, these groups of two are given by

\begin{minipage}{.45\linewidth}
\begin{alignat*}{2}
    \Box& =&{} - b_{1} b_{2} \Box \\
    \e{1} \Box& =&{} - \ebar{1} b_{2} \Box \\
    \e{2} \Box& =&{} - b_{1} \ebar{2} \Box \\
    \e{1}\e{2} \Box& =&{} - \ebar{1}\ebar{2} \Box
\end{alignat*}
\end{minipage}%
\begin{minipage}{.45\linewidth}
\begin{align*}
    b_{1}\Box &= b_{2} \Box \\
    \ebar{1} \Box &= \e{1} b_{2} \Box \\
    \ebar{2} \Box &= \e{2} b_{1} \Box \\
    \ebar{1}\e{2} \Box &= \e{1}\ebar{2} \Box \ .
\end{align*}
\end{minipage}

\noindent
Thus the forgetful idempotent defines a left-ideal, and the elements of this ideal are called \emph{spinors}. 
The $2^{3}$ dimensional \emph{real} spinor space is defined as the left ideal generated by $\Box$:
    \begin{equation}\label{eq:spinor_space}
        S =  \{ X \Box \; \vert \; X \in \GR{4} \} \ .
    \end{equation}
Moreover, we can distinguish the even and odd spinor spaces $S^\pm$ (note that $\Box$ is an even grade element):
    \begin{equation}
        S^\pm =  \{ X \Box \; \vert \; X \in \GR[\pm]{4} \} \ .
    \end{equation}
Both of these are $2^2$ real dimensional. 
However, the above definition of $S$ features a two-to-one mapping from \GR{4} to $S$. A more convenient and equivalent one-to-one mapping is obtained if we define $S$ as
    \begin{equation}\label{eqn_R3}
        S \coloneqq \{ X \Box  \; \vert \; X \in \GR{3} \} \ ,
    \end{equation}
where $\GR{3} = \text{Alg}(\e{1}, \e{2}, \ebar{1})$ is the algebra generated by $\{ \e{1}, \e{2}, \ebar{1} \}$. Note that we picked up one reflection in each of the points $b_j$ appearing in the spine $\rrho_0$. However, this choice is not unique; we could have picked any reflection through each point $b_j$. In order to account for this, we must choose one (and {\em only} one) additional orthogonal reflection.
To see why the algebra $\GR{3}$ suffices, note that any element $X \in \GR{4}$ can be written as
\[ X = X_0 + X_1\ebar{1} + X_2\ebar{2} + X_{12}\ebar{1}\ebar{2}\ , \]
where each $X_a \in \GR{2}$ belongs to the algebra generated by $\e{1}$ and $\e{2}$. Multiplying $X$ with $\Box$ from the right, we can use the fact that 
\begin{align*}
\ebar{2}\Box &= \e{2}b_2\Box = \e{2}b_1\Box = \e{2}\e{1}\ebar{1}\Box \\
\ebar{1}\ebar{2}\Box &= \ebar{1}\e{2}^2\ebar{2}\Box = \ebar{1}\e{2}b_1\Box = \e{2}\e{1}\Box
\end{align*} 
to conclude that $X\Box$ indeed belongs to $\GR{3}\Box$ as defined above. Going one step further, we can now say that 
\[ \GR{3} = \GR{2} \oplus \GR{2}\ebar{1} = \GR{2} \otimes \big(\mathbb{R} \oplus \mathbb{R}\e{1}\ebar{1}\big) \cong \GR{2} \otimes \mathbb{C} \cong \GR{2} \otimes \GR{0,1}\ . \]
Note that the `complex number' $z = x + y b_1$ commutes with the box, which means that every spinor $\Psi = X\Box$ can thus be written as 
$\Psi = A \Box + B \Box \bl_1$, where the complex phase factor appears at the right-hand side and with $A, B \in \GR{2}$. 
We can therefore alternatively write $S$ as
    \begin{equation}
        S \coloneqq \{ A \Box + B \Box \bl_1  \; \vert \; A, B \in \GR{2} \} \ ,
    \end{equation}
Using this mapping we can also define a grading on $S$ inherited from the natural grading on \GR{2}, which we thus define as
    \begin{equation}
        S^{(j)} \coloneqq \{ A \Box + B \Box \bl_1  \; \vert \; A, B \in \GR[(j)]{2} \} \ .
    \end{equation}
In total, we thus arrive at $4 \times 2 = 8$ real DOF. This is in accordance with the 4 complex DOF in a classical approach, since the spinor representation for $\Spin{4} \cong \SU{2} \times \SU{2}$ is of the form $\mathbb{C}^2 \otimes \mathbb{C}^2$.
Hence, at least based on this counting argument, $S$ has the right number DOF to be the spinor space of \Spin{4}, despite the fact that we did not have to introduce complex numbers.
Amazingly, in the defining matrix representation of \SU{2} this rich structure was hiding under the null-matrix all along. 
We will now proceed to justify why it is correct to call $S$ the spinor space.

\section{Spinors}\label{sec:spineors}

Recall that the spine $\rrho_0$ has \UN{2} as its stabilizer group, and hence 
    \[ U \rrho_0 \widetilde{U} = \rrho_0 \]
for any $U \in \UN{2}$.
We then asked for a way to lose the distinction between the commuting imaginary units $\bl_j$, and found the forgetful idempotent $\Box = 1 - \I$.
But from this definition of the forgetful idempotent it immediately follows that $\rrho_i \Box = \Box \rrho_i = 0$ for $i = 1, 2, 3$, and so $\Box$ 
is invariant even under single-sided multiplication by $U \in \SU{2}$ elements:
\[ U \Box \widetilde{U} = U \Box = \Box \widetilde{U} = \Box \ . \]
Since pure spinors in $d = 2k$ dimensions are known to be invariant under single-sided multiplication by elements of \SU{k} \cite{Charlton1997}, the forgetful idempotent is the real version of a pure spinor.

Since general elements $\Psi$ of the spinor space $S$ are 
of the form $\Psi = X \Box$ with $X \in \GR{3}$, we find that arbitrary spinors transform under $U \in \SU{2}$ as
    \[ U \Psi \widetilde{U} = U X \Box \widetilde{U} = (U X) \Box = U \Psi. \]
Notice how \emph{spinors transform under two-sided conjugation} just like everything else, but they simply forget the effect of $\widetilde{U}$ coming from the right.
Similarly the conjugate spinor $\widetilde{\Psi}$ transforms as
    \[ U \widetilde{\Psi} \widetilde{U} = U \Box \widetilde{X} \widetilde{U} = \Box (\widetilde{X} \widetilde{U}) = \widetilde{\Psi} \widetilde{U}, \]
and has completely forgotten the effects of $U$ coming from the left.

However, if $U = \exp(\theta \rrho_0)$ is an isoclinic rotation generated by the spine $\rrho_0$ itself and hence an element of $\UN{1}$, 
then $\Box$ commutes with $U$ but does not forget it, and we find
    \[ U \Psi \widetilde{U} = (U X \widetilde{U}) \Box. \]
This highlights that the forgetfulness of $\Box$ is very selective: it only occurs for $U \in \SU{2}$, but not for general $R \in \Spin{4}$.
So we stress that spinors transform two-sidedly just like everything else, but they might forget about part of the transformation, giving the same effect as a single-sided transformation.
But we will only perform two-sided transformations from here on out: \emph{it is up to the elements to transform as they will}.

\subsection*{Classifying the elements of \texorpdfstring{$S$}{S}}\label{sec:classification}
We will now proceed to classify the elements of the spinor space $S$, by studying how the elements of $S$ transform under \UN{2} transformations.
For any unitary transformation $U = \exp(\theta \rrho / 2) \in \UN{2}$ with $\rrho \in \un{2}$ we find at first order
    \begin{align}
        U \Psi \widetilde{U} &= \Psi + \tfrac{1}{2} \theta (\rrho \Psi - \Psi \rrho) + \order{\theta^2} \\
        &\approx \Psi + \theta \rrho \times \Psi, \notag
    \end{align}
where $\rrho \times \Psi \coloneqq \tfrac{1}{2} (\rrho \Psi - \Psi \rrho)$ is the commutator product.

The elements of $S$ can be classified by their eigenvalues under the commutator product 
with each of the basis elements of the Cartan subalgebra $\mathfrak{h}$ of \un{2} 
given by the elements
    \[ \mathfrak{h} = \text{span}\{ \rrho_0, \rrho_3 \}.\]
Measuring eigenvalues is a delicate matter that needs to be handled with some care.\footnote{Traditional texts start from the eigenvectors $v_{\pm j} = \tfrac{1}{2}(\e{j} \pm i \ebar{j})$ of the simple bivectors $\bl_j = i \Sigma_j$, satisfying $\bl_j v_{\pm j} = \pm i v_{\pm j}$ to define a (pure) spinor $\boxplus = \prod_{j=1}^k v_{+j}v_{-j}$ that is a $+i$ eigenstate for all $\bl_j$ since $\bl_j \boxplus = +i \boxplus$ (hence our notation "plus box"). But this eigenequation literally says that $\boxplus$ is a forgetful idempotent imposing an equivalence between $i$ and $\bl_j$, and so complexification does not add any new information that could not already be expressed by the $\bl_j$. In fact, $\Box \propto \Re(\boxplus)$,
and so it is $\Box$ that captures all the spinorial behaviour, and as this paper shows, $\Box$ can be formulated directly from the $\bl_j$.
However, there is one big advantage to using $i$ that we have sacrificed by telling the real story: $i$ commutes with the entire algebra and so the eigenequation is always $B \times \chi = \lambda i \chi$ for any $\chi \in \{ S, \widetilde S \}$, whereas we always need to ensure that $\bl$ is next to $\Box$ in order to perform a correct measurement of the eigenvalue.
}
Note how each element $\B \in \mathfrak{h}$ of the Cartan subalgebra $\mathfrak{h}$ becomes a scalar multiple of the $\bl_j$ once it is thrown into $\Box$, since they are all linear combinations of the $\bl_j$.
In order to measure the correct sign of the eigenvalues, it is therefore important to have $\bl_j$ directly adjecent to $\Box$.
Hence, the eigenequation we need to solve is
    \begin{equation}\label{eq:eigenequation}
        \B \times \Psi = \lambda \Psi b, 
    \end{equation}
where $\lambda$ is the eigenvalue of $\Psi$ under $B$ and $b$ is either $b_1$ or $b_2$. 
Following long standing physics conventions, we define the operators for hypercharge, isospin, and electric charge as
    \[ \tfrac{1}{2} Y = \rrho_0, \qquad  I_3 = \rrho_3, \qquad Q = I_3 + \tfrac{1}{2}Y = \rrho_3 + \rrho_0 = b_1. \]
As an example, let us calculate the hypercharge of $\Psi = \e{1} \Box$:
    \begin{align*}
        2 \rrho_0 \times \Psi &= 2 (\rrho_0 \times \e{1}) \Box  = (\bl_1 \times \e{1} + \cancel{\bl_2 \times \e{1}}) \Box \\
        &=  \bl_1 \e{1} \Box = -  \e{1} \bl_1 \Box = -  \Psi \bl \ ,
    \end{align*}
where we have used $\bl_1 \times \e{1} = \bl_1 \cdot \e{1} = \bl_1 \e{1}$. 
Thus we find that $\lambda = - 1$, which is the hypercharge of a left-handed lepton.
Using similar computations the eigenvalues of all elements of $S$ under the bivectors $\B \in \{ Y, I_3\}$ can be computed. 
In addition the eigenvalues of $\widetilde{\Psi} \in \widetilde{S}$ can be computed as $\B \times \widetilde{\Psi} = \lambda b \widetilde{\Psi}$, once again ensuring that $b$ is directly adjacent to $\Box$.
The eigenvalues of $S$ and $\widetilde{S}$ under $\mathfrak{h}$ are shown in \cref{tab:quantum_numbers}.

    \begin{table}[tbp]
    \centering
    \renewcommand{\arraystretch}{1.3}
    \begin{tabular}{
      l|
      >{\columncolor{Scolor}}r
      >{\columncolor{Scolor}}r
      >{\columncolor{Scolor}}r
      >{\columncolor{Scolor}}r|
      l|
      >{\columncolor{Stildecolor}}r
      >{\columncolor{Stildecolor}}r
      >{\columncolor{Stildecolor}}r
      >{\columncolor{Stildecolor}}r|
      l}
    
    \hhline{~|====|~|====|}
    \rowcolor{gray!15}
    \cellcolor{white}{} & \multicolumn{4}{c|}{\cellcolor{Scolor}\textbf{$S$}} & \multicolumn{1}{c|}{\cellcolor{white}{}} & \multicolumn{4}{c|}{\cellcolor{Stildecolor}\textbf{$\widetilde{S}$}} & \multicolumn{1}{c}{\cellcolor{white}{}} \\
    \cellcolor{white}{} & \multicolumn{1}{c}{\cellcolor{Scolor}$\tfrac{1}{2} Y$} & \multicolumn{1}{c}{\cellcolor{Scolor}$I_3$} & \multicolumn{1}{c}{\cellcolor{Scolor}$Q$} & \multicolumn{1}{c|}{\cellcolor{Scolor}{}} & \multicolumn{1}{c|}{\cellcolor{white}{}} & \multicolumn{1}{c}{\cellcolor{Stildecolor}$\tfrac{1}{2} Y$} & \multicolumn{1}{c}{\cellcolor{Stildecolor}$I_3$} & \multicolumn{1}{c}{\cellcolor{Stildecolor}$Q$} & \multicolumn{1}{c|}{\cellcolor{Stildecolor}{}} & \multicolumn{1}{c}{\cellcolor{white}{}} \\
    
    \hhline{~|====|=|====|}
    
    \rowcolor{headercolor}
    \multicolumn{1}{c|}{\cellcolor{white}{\textbf{id}}} & $\rrho_0$ & $\rrho_3$ & $\bl_1$ & $\bl_2$ & \multicolumn{1}{c|}{\cellcolor{Xcolor}$X$} & $\rrho_0$ & $\rrho_3$ & $\bl_1$ & $\bl_2$ & \multicolumn{1}{l}{\cellcolor{white}{\textbf{id}}} \\
    \cline{2-6} \cline{7-10}
    
    \rowcolor{nuRcolor}
    \multicolumn{1}{c|}{\cellcolor{white}{\textcolor[HTML]{6EA1D4}{$\nu_R$}}} & $0$ & $0$ & $0$ & $0$ & \multicolumn{1}{c|}{\cellcolor{Xcolor}$z$} &
    $0$ & $0$ & $0$ & $0$ & \multicolumn{1}{l}{\cellcolor{white}{\textcolor[HTML]{6EA1D4}{$\overline{\nu}_L$}}} \\
    
    \rowcolor{epsilonLcolor}
    \multicolumn{1}{c|}{\cellcolor{white}{\textcolor[HTML]{CE3375}{$\varepsilon_L$}}} & $-\tfrac{1}{2}$ & $-\tfrac{1}{2}$ & $-1$ & $0$ & \multicolumn{1}{c|}{\cellcolor{Xcolor}$\e{1}z$} &
    $\tfrac{1}{2}$ & $\tfrac{1}{2}$ & $1$ & $0$ & \multicolumn{1}{l}{\cellcolor{white}{\textcolor[HTML]{CE3375}{$\overline{\varepsilon}_R$, $\phi^+$}}} \\

    \rowcolor{nuLcolor}
    \multicolumn{1}{c|}{\cellcolor{white}{\textcolor[HTML]{E881A6}{$\nu_L$}}} & $-\tfrac{1}{2}$ & $\tfrac{1}{2}$ & $0$ & $-1$ & \multicolumn{1}{c|}{\cellcolor{Xcolor}$\e{2}z$} &
    $\tfrac{1}{2}$ & $-\tfrac{1}{2}$ & $0$ & $1$ & \multicolumn{1}{l}{\cellcolor{white}{\textcolor[HTML]{E881A6}{$\overline{\nu}_R$, $\phi^0$}}} \\
    
    \rowcolor{epsilonRcolor}
    \multicolumn{1}{c|}{\cellcolor{white}{\textcolor[HTML]{1B5091}{$\varepsilon_R$}}} & $-1$ & $0$ & $-1$ & $-1$ & \multicolumn{1}{c|}{\cellcolor{Xcolor}$\e{12}z$} &
    $1$ & $0$ & $1$ & $1$ & \multicolumn{1}{l}{\cellcolor{white}{\textcolor[HTML]{1B5091}{$\overline{\varepsilon}_L$}}} \\
    \cline{2-6} \cline{7-10}
    
    \end{tabular}
    \caption{The eigenvalues of $\Psi = X \Box$ for various $X$ under the commutator product with the operators $\tfrac{1}{2}Y = \rrho_0$, $I_3 = \rrho_3$ and $Q = I_3 + \tfrac{1}{2} Y = \bl_1$, where $z = x + y \bl_1$ is a `complex number'. The eigenvalues of $\Psi$ under $\bl_2$ are included for completeness, although $\bl_2$ does not correspond to known physics. The first and last columns show the identification with Standard Model particles. For elements in $S$ the eigenvalues are computed as $\rrho \times \Psi = \lambda \Psi b $, whereas for elements of $\widetilde{S}$ they are computed as  $\rrho \times \widetilde{\Psi} = \lambda b \widetilde{\Psi} $, see text for the justification.}
    \label{tab:quantum_numbers}
    \end{table}

The pattern that emerges is very encouraging. 
The eigenvalues under $I_3$ tell us that $S^{(0)}$ and $S^{(2)}$ transform as isospin singlets, while $S^{(1)}$ transforms like an isospin doublet. 
This implies that left-handed leptons should be elements of $S^{(1)}$ since they couple to isospin \SU{2}, while right-handed leptons should be elements of $S^{(0)}$ and $S^{(2)}$ because they do not interact with isospin.
On the basis of the eigenvalues under $Y$ and $Q$ we see that we must identify $S^{(0)}$ with the right-handed neutrino $\nu_R$, while $S^{(2)}$ must be a right-handed electron $\varepsilon_R$.
We then assign the labels $\nu_L$ and $\varepsilon_L$ to the elements of $S^{(1)}$ such that an electron $\varepsilon$ always has electric charge $-1$, while a neutrino $\nu$ must always be electrically neutral.

Moreover, we already know that while $\Psi$ transforms under $U$ in $\SU{2}$ as $\Psi \to U \Psi$, the reversed spinor $\widetilde{\Psi}$ transforms as $\widetilde{\Psi} \to \widetilde{\Psi} \widetilde{U}$, and hence $\widetilde{S}$ is home to the corresponding anti-particles.
After all, the spaces ${S}^{(1)}$ and $\widetilde{S}^{(1)}$ are incompatible \SU{2} representations since it is impossible to write e.g. $\Box \e{1}$ as a left ideal using $\Box$ since
    \[ \Box \e{1} = \e{1} \I  \not\propto \Box \ .\]
So based on the $Y$, $I_3$, and $Q$ eigenvalues, the obvious interpretation of the elements of $\widetilde{S}$ is as anti-particles.
However, since $\widetilde{S}^-$ has states with eigenvalues $Y = 1$ and $I_3 = \pm \tfrac{1}{2}$, it might also be home to the $\phi^+$ and $\phi^0$ components of the Higgs field!
This will be investigated further in a follow-up paper.
A short remark of the assignment of labels to the elements of $\widetilde{S}$ as given in \cref{tab:quantum_numbers} is required. In order to match the fact that the charge conjugate of a right-handed particle is a left-handed anti-particle, we assign e.g. $\widetilde{\epsilon}_R = \overline{\epsilon}_L$, where an overline is traditionally used to denote an anti-particle, since the eigenvalues force us to do so in order to keep our notation consistent with other texts.\footnote{The attentive reader might protest that $\widetilde{\epsilon}_R = \overline{\epsilon}_L$ is not consistent with our claim in the following paragraph that grade-parity is to be interpreted as chirality, since reversion is grade-preserving. Bear in mind however, that in the current panel we are only focusing on the gauge sector while ignoring the spacetime part, and clearly the eigenvalues force us to make this identification of irreps to particle labels. 
Hence, the complete charge conjugation operation must do more than just reversion in the spacetime part, but this will be a topic for the next panels in the triptych.}

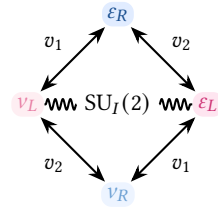
\begin{figure}[t]
    \centering
    \begin{tikzpicture}[>=Stealth,scale=1.2]

        \node[fill=nuLcolor,rounded corners,inner sep=2pt,draw=none] (nuL) at (-1,0) {\textcolor[HTML]{E881A6}{$\nu_L$}};
        \node[fill=epsilonRcolor,rounded corners,inner sep=2pt,draw=none] (eR)  at (0,1) {\textcolor[HTML]{1B5091}{$\varepsilon_R$}};
        \node[fill=nuRcolor,rounded corners,inner sep=2pt,draw=none] (nuR) at (0,-1) {\textcolor[HTML]{6EA1D4}{$\nu_R$}};
        \node[fill=epsilonLcolor,rounded corners,inner sep=2pt,draw=none] (eL)  at (1,0) {\textcolor[HTML]{CE3375}{$\varepsilon_L$}};
        
        \draw[<->,thick,black] (nuL) -- node[above left,text=black] {$v_1$} (eR);
        \draw[<->,thick,black] (eR) -- node[above right,text=black] {$v_2$} (eL);
        \draw[<->,thick,black] (eL) -- node[below right,text=black] {$v_1$} (nuR);
        \draw[<->,thick,black] (nuR) -- node[below left,text=black] {$v_2$} (nuL);
        
        \node (su2) at (0,0) {$\SU[I]{2}$};
    
        \draw[
          decorate,
          decoration={snake, amplitude=0.7mm, segment length=1.0mm},
          thick,
        ] (su2.west) -- (nuL.east);
    
        \draw[
          decorate,
          decoration={snake, amplitude=0.7mm, segment length=1.0mm},
          thick,
        ] (su2.east) -- (eL.west);
    
    \end{tikzpicture}
    \caption{Fock space structure of the particle states. Note how the arrows are bi-directional, since the vectors $v_i = e^{\theta_i \bl_i} \e{i}$ serve as both raising and lowering operators, captured algebraically by the fact that $v_{i}^2 = 1$.
    Moreover, the $\nu_L$ and $\varepsilon_L$ states couple to \SU[I]{2}.
    }
    \label{fig:fock}
\end{figure}

\Cref{fig:fock} presents the results of \cref{tab:quantum_numbers} in the form of a Fock space diagram. Let $v_i = e^{\theta_i \bl_i} \e{i} = \cos\theta_i \e{i} + \sin\theta_i \ebar{i}$ be a unit vector satisfying $\bl_i \wedge v_i = 0$ and $v_i^2 = 1$. Then the particle states $\nu_{L}$, $\varepsilon_{L}$ and $\varepsilon_{R}$ can be reached from the ground state $\nu_R = z \Box$ by multiplying by $v_1$, $v_2$ and $v_1 v_2$, respectively. The vectors $v_i$ therefore serve as both raising and lowering operators, which is captured by the fact that $v_i^2 = 1$. Rather than calling them ladder operators, it might be more appropriate to refer to them as toggle operators.

Promisingly, it appears as though \GR{4} naturally houses chiral leptons, where left-handed particles and right-handed anti-particles participate in isospin \SU{2}, while right-handed particles and left-handed anti-particles do not participate in this interaction. 
This is a very promising sign, because it means we have found a purely geometric reason for why left- and right-handed particles behave differently under isospin: \underline{chirality corresponds to grade-parity}.

But to be able to make this claim beyond a shadow of a doubt, we must ensure 
that different choices of spine will always lead one to this conclusion. Ignoring a change of basis, which obviously does not affect our claim, there is a second spine to consider:
$\rrho_3$ is also a spine, orthogonal to $\rrho_0$ since $\rrho_0 \rrho_3 = 0$, and so there is a second \SU{2} group which leaves $\rrho_3$ invariant rather than $\rrho_0$.
Thus, if Alice had chosen $\rrho_0$ as her spine, while Bob had chosen $\rrho_3$ as his, would they both agree that chirality corresponds to grade-parity?

In order to investigate this let us introduce suitable notations for the two copies of \SU{2} which are in play here. Isospin \SU{2}, which will be denoted by \SU[I]{2} from now on, is the image under the exponential map of its Lie algebra $\mathfrak{su}_I(2)$ generated by the Pauli bivectors \cref{eq:paulibivectors} which can be written as
\[ \rrho_1 = \ebar{1}\ebar{2} \I, \quad \rrho_2 = \e{1}\ebar{2} \I, \quad \rrho_3 = -\e{2}\ebar{2} \I\ . \]
On the other hand, \emph{hyperspin} \SU{2}, denoted by \SU[Y]{2}, is the invariance group of the orthogonal spine $\rrho_3$, for which the hypercharge operator $\rrho_0$ then acts as its $i \sigma_3$. This group is obtained as the image of the Lie algebra $\mathfrak{su}_Y(2)$ generated by its own set of (Pauli) bivectors 
\[  \rrho_{-1} = \ebar{1}\ebar{2} \Box, \quad \rrho_{-2} = \e{1}\ebar{2} \Box, \quad \rrho_0 = -\e{2}\ebar{2} \Box \ .  \]
Note that arbitrary elements in $\mathfrak{su}_I(2)$ and $\mathfrak{su}_Y(2)$ are thus of the form $B = v\ebar{2}\mathcal{I}$, where $v = \alpha \e{1} + \beta \e{2} + \gamma \ebar{1} \in \mathbb{R}^3$ (with $\alpha, \beta, \gamma \in \mathbb{R}$ and $\mathbb{R}^3$ as in \cref{eqn_R3} above) and $\mathcal{I}$ an idempotent which then respectively stands for $\I$ or $\Box$. As a result, it follows from a standard Taylor expansion that
\begin{align}\label{formula_exp_su2}
    \exp(v\ebar{2}\mathcal{I}) &= (1 - \mathcal{I}) + (\cos|v| + \hat{v}\ebar{2}\sin|v|)\mathcal{I}\nonumber\\
    &= (1 - \mathcal{I}) + \exp(v\ebar{2})\mathcal{I}\ .
\end{align}
with $v = |v|\hat{v}$. It is worth noting here that for $\mathcal{I} = \I$ one will get $1 - \mathcal{I} = \Box$ and vice versa. So $\exp(B\mathcal{I})$ always contains both $\Box$ and $\I$. In a sense, this is still true when $B = 0$, because $\exp(0) = 1 = \I + \Box$. Also note that for any bivector $B \in \GR[2]{4}$ one has that 
\[ B = B\I + B\Box\ \Rightarrow\ s := \exp(B) = \exp(B\I)\exp(B\Box)\ , \]
which realises every element $s$ in Spin$(4)$ as a product $s = U_IU_Y$ of an isospin and a hyperspin element. This last factor explains why spinors {\em only} seem to transform one-sidedly when $U_Y = 1$ is trivial, with $U_I\Psi \widetilde{U}_I = U_I\Psi$ for $\Psi = X\Box$ in that case. In general, one has that 
\[ s\Psi\widetilde{s} = U_I\big(U_Y\Psi\widetilde{U}_Y\big)\ . \]
The upshot of \cref{formula_exp_su2} is that when acting on spinors in $S$, either $(1 - \mathcal{I})$ or $\exp(e_1v)\mathcal{I}$ will act trivially (depending on the grade of the spinor) and this fixes the transformation behaviour (see the next section for a more detailed explanation, when we will consider the spinor products). This thus implies that even and odd spinors will always behave differently under the (left multiplication\footnote{Note that we have to be specific about the one-sided action here, since $U_Y \in \SU[Y]{2}$ does not act as the identity element on $\Psi = X\Box$ when acting from the right.}) action of the group \SU{2}, be it isospin or hyperspin. As a matter of fact, we get the following table: 
    \begin{table}[!h]
        \centering
        \begin{tabular}{l|c|c|}
            \cline{2-3}
            & \SU[I]{2}  & \SU[Y]{2}\\
            \hhline{~==}
            $S^+$ & \multicolumn{1}{c|}{\cellcolor{nuRcolor}$\mathbb{C} \oplus \mathbb{C}$} & \multicolumn{1}{c|}{\cellcolor{nuLcolor}$\mathbb{C}^2$}\\
            \cline{2-3}
            $S^-$ & \multicolumn{1}{c|}{\cellcolor{nuLcolor}$\mathbb{C}^2$} & \multicolumn{1}{c|}{\cellcolor{nuRcolor}$\mathbb{C} \oplus \mathbb{C}$} \\
            \cline{2-3}
        \end{tabular}
    \end{table}
Similar conclusions also hold for the (right) action on $\widetilde{S}$. This table clearly shows that the distinction between isospin and hyperspin does not respect the grades: even spinors in $S^+$ are singlets under the former, but behave as a doublet under the latter, and vice versa for odd spinors. 
However, while Alice and Bob might not agree on which elements transform as what under their respective $\SU{2}$, they will agree grade-parity is what determines the transformation behavior, and they could agree to call this grade-parity chirality.

Alternatively, there is a deeper interpretation possible if one takes the orientation of the origin into account. We refer to \cite{DKER2024} for the motivation behind this somewhat controversial idea, because points are mostly thought of as boring zero-dimensional entities bearing no (internal) structure at all. But in the plane-based view for $\GR{4}$, the origin $O$ can be identified with a multiple of the pseudoscalar $\e{1}\ebar{1}\e{2}\ebar{2}$. 
Indeed, since the generators represent hyperplanes in $\mathbb{R}^4$, their product corresponds to the common intersection. After the normalisation $O\widetilde{O} = 1$, there are essentially 2 possible orientations left, with $O = \pm b_1b_2$. 
A transformation $R \in \Spin{4}$ preserves the orientation of the origin under conjugation, whereas an odd transformation $P \in \Pin[-]{4}$ inverts the orientation.
One could therefore consider tranformations of the type $c_+ R + c_- P$ acting on the origin, which we have dubbed \emph{pointors} in the paper \cite{DKER2024}.
This clearly shows that the grade of the transformation is connected to the orientation of the origin.
This behaviour not only manifests itself on the level of the origin, but even on the level of the underlying spine. To see this, we first note that $+O = 4 \rrho_0 \wedge \rrho_0$ whereas $-O = 4 \rrho_3 \wedge \rrho_3$. The upshot here is that $\rrho_0$ and $\rrho_3$ can also be transformed into each other using a single reflection, with for instance $\ebar{2}\rrho_0\ebar{2}^{-1} = \rrho_3$. Because the spine (be it $\rrho_0$ or $\rrho_3$) dictates which $\SU{2}$ group one is using (isospin or hyperspin), this essentially means that the transition from isospin to hyperspin comes from a parity switch. 
Therefore, regardless of the choice one makes for the spine (whereby our choice for $\rrho_0$ lies closest to long-standing traditions in physics), even and odd spinors will always behave differently as a representation space for $\SU{2}$.

\section{Spinor Products}\label{sec:spinorproducts}
There are two products between spinors $\Psi, \Phi \in S$ that need to be considered: the innor product $\widetilde{\Psi} \Phi $ and the outor product $\Psi \widetilde{\Phi}$.
The innor product plays an important role when calculating magnitudes, and therefore features heavily in the Standard Model Langrangian.
Meanwhile, the outer product appears in particle interactions.

\subsection{Innor product}\label{sec:innorproducts}

Traditional Pauli spinors for $\SU{2}$ are of the form
$$ \Psi = \begin{pmatrix} \psi_1 \\ \psi_2 \end{pmatrix} \quad \text{where}\quad \psi_1, \psi_2 \in \mathbb{C} \ .$$
There are two scalar products between Pauli spinors \cite{lounesto}:
\begin{alignat*}{2}
L_1 &:= \Psi^\dagger \Phi &= \psi_1^* \varphi_1 + \psi_2^* \varphi_2 \\
L_2 &:= \Psi^T i \sigma_2 \Phi &= \psi_1 \varphi_2 - \psi_2 \varphi_1   
\end{alignat*}
where $z^*$ stands for the complex conjugate. The product $L_1$ is the traditional Hermitian inner product, written in bra-ket notation as $\braket{\Psi}{\Phi}$, while the product $L_2$ is the symplectic inner product.
Now consider writing both $\Psi$ and $\Phi$ as elements of $S^+$ instead of as column vectors. We have
\begin{align*}
    \Psi &= {\psi} \Box = (\psi_{1} + \mathbf{e}_{12} \psi_2 ) \Box, \quad \Phi = \varphi \Box = (\varphi_{1} + \mathbf{e}_{12} \varphi_2 ) \Box,
\end{align*}
where $\psi_{i}$ and $\varphi_{i}$ are "complex" numbers of the form $\alpha + \beta b$. 
Notice that the coefficients $\psi \coloneqq \psi_{1} + \mathbf{e}_{12} \psi_2$ and $\varphi \coloneqq \varphi_{1} + \mathbf{e}_{12} \varphi_2$ are simply quaternions in $\mathbb{R}_3 = \text{Alg}(\e{1}, \e{2}, \ebar{1})$.
Then 
\begin{align*}
    \widetilde{\Psi} \Phi &= \Box (\widetilde{\psi}_{1} + \widetilde{\psi}_2 \mathbf{e}_{21}) (\varphi_{1} + \mathbf{e}_{12} \varphi_2 ) \Box \\
    &= \bqty{\widetilde{\psi}_{1} \varphi_{1} + \widetilde{\psi}_2 \mathbf{e}_{21} \varphi_{1} + \widetilde{\psi}_{1} \mathbf{e}_{12} \varphi_2 + \widetilde{\psi}_2 \varphi_2 } \Box \\
    &= \bqty{\widetilde{\psi}_{1} \varphi_{1} + \widetilde{\psi}_2 \varphi_2 + \mathbf{e}_{12} ({\psi}_{1}  \varphi_2 - {\psi}_2 \varphi_{1}) } \Box  \ ,
\end{align*}
where we have used $S^+ \Box = \Box S^+$ and $\widetilde{\psi}_i \e{12} = \e{12} \psi_i$.
The $L_1$ and $L_2$ norms can be retrieved as
    \begin{align*}
        L_1 = \langle\widetilde{\Psi} \Phi \rangle, \quad L_2 = \langle \e{21} \widetilde{\Psi} \Phi \rangle \ ,
    \end{align*}
where $\expval{\ldots}$ is grade projection onto the scalar part.
Hence the product $\widetilde{\Psi} \Phi$ represents a novel spinor product
 \[ \Psi * \Phi = \begin{pmatrix}
     \psi_1^* \varphi_1 + \psi_2^*\varphi_2, & \psi_1 \varphi_2 - \psi_2 \varphi_1    
 \end{pmatrix} \ , \]
which computes the products $L_1$ and $L_2$ simultaneously,
and thus the product $\widetilde{\Psi} \Phi$ presents an important unification of two previously unconnected products, somewhat reminiscent of the unification of the inner and exterior product in the geometric product. Moreover, the product $\widetilde{\Psi} \Phi = \widetilde{\psi} \varphi \Box$ is itself again an element of $S^+$, since $\widetilde{\psi} \varphi$ is a quaternion in \GR{3}.
Hence, this product between spinors deserves its own name: the \emph{innor product}.

Moreover, in the previous discussion $\Psi$ and $\Phi$ were assumed to be even (right-handed). But odd (left-handed) spinors $\Psi_L = v \psi \Box$ and $\Phi_L = v \varphi \Box$ with $v \in \GR[3]{}$ any reference unit vector have the same innor product, since 
\[ \widetilde{\Psi}_L \Phi_L = \Box \widetilde{\psi} v v \varphi \Box = \widetilde{\psi} \varphi \Box \ . \]
However, the innor product between a left-handed (odd) spinor $\Psi_L$ and a right-handed (even) spinor $\Phi_R$ annihilates, since
\[ \widetilde{\Psi}_L \Phi_R = \Box \widetilde{\psi} v \varphi \Box = \Box \I \widetilde{\psi} v \varphi = 0 \ . \]
Subsequently, the innor product maps either two left-handed (odd) or two right-handed (even) spinors to a right-handed (even) spinor, while the product between right and left (even and odd) vanishes: 
    \begin{align}
        \widetilde{\Psi}_L \Phi_L \in S^+, \quad \widetilde{\Psi}_R \Phi_R \in S^+, \quad \widetilde{\Psi}_L \Phi_R = \widetilde{\Psi}_R \Phi_L = 0 \ .
    \end{align}
Alternatively the fact that the product must be an element of $S^+$ follows directly because the innor product is manifestly invariant under transformations $U \in \SU{2}$:
    \[ \widetilde{\Psi} \Phi \to (U \widetilde{\Psi} \widetilde{U}) (U \Phi \widetilde{U}) = (\widetilde{\Psi} \widetilde{U}) (U \Phi) = \widetilde{\Psi} \Phi, \]
and hence $\widetilde{\Psi}\Phi$ is an isospin singlet, and thus an element of $S^+$. (As the adage goes, a singlet is something that transforms like a singlet.)

The innor product can also be used to define the (squared) norm of a spinor. Indeed, since
$\widetilde{\Psi} \Psi = \alpha\Box$ with $\alpha \in \mathbb{R}_{\geq 0}$, it suffices to put 
\[ \|\Psi\|^2 := 2\expval{\widetilde{\Psi} \Psi} \ , \] 
where $\expval{\ldots}$ selects the scalar part.
That $\widetilde{\Psi} \Psi$ must be a scalar multiple of $\Box$ follows from the fact that this product is self-reverse, and $\Box$ is the only self-reverse element in $S^+$. Moreover, this scalar $\alpha$ is always positive (strictly positive for $\Psi \neq 0$) because $\expval{\widetilde{X}X} = \sum_A X_A^2$ for any $X \in \GR{4}$, where $X_A \in \mathbb{R}$ denotes the coefficients with respect to the standard basis. 

Getting slightly ahead of ourselves, the properties of the innor product will play an important role when we build the Yukawa interactions between the leptons $\ell \in S$ and the Higgs field $\Phi$.
After all, the innor product prevents us from adding a lepton mass term
    \[ \Delta \mathcal{L} = m( \widetilde{\ell}_L \ell_R + \widetilde{\ell}_R \ell_L ) = 0 \]
to the Lagrangian since it is trivially zero.
However, it also offers an elegant way out, since we can add terms
    \[ \Delta \mathcal{L} = \widetilde{\ell}_R (\widetilde{\Phi} \ell_L)  + (\widetilde{\ell}_L \Phi) \ell_R \ ,\]
which can only be non-zero if $\Phi$ itself is also an isospin doublet, and hence odd. 
Since the terms $\widetilde{\ell}_R (\widetilde{\Phi} \ell_L)$ and $(\widetilde{\ell}_L \Phi) \ell_R$ are each others reverse, such a term is proportional to $\Box$, as it must be.
This is precisely the way the Higgs gives mass to fermions in the Standard Model Lagrangian.

As is evident from the spinor norm and the Higgs coupling above, the innor product will play an important role in constructing the Standard Model Lagrangian.
However, there is another product that plays an important role in particle interactions: the \emph{outor product}.

\subsection{Outor products}\label{sec:outerproducts}

In this section we will consider spinor products of the form $\Psi \widetilde{\Phi}$ for $\Psi, \Phi \in S$. 
Such products are important in e.g. the computations of fermionic correlation functions or Fermi's theory of the electroweak interaction.
We will now proceed to decompose the product $\Psi \widetilde{\Phi}$ into irreps, depending on the grade (or irrep) of the input, to establish a link back to the representation theory of \SU{2}.
But it must be stressed that the main take-away of this section is that there is no need to get down to the level of irreps: we can work with $\Psi \widetilde{\Phi}$ directly.

\subsubsection*{Product between isospin doublets}
Let us consider the interaction between two left-handed leptons $\Psi_L, \Phi_L \in S^{(1)}$, in the product $\Psi_L \widetilde{\Phi}_L$. 
Since $\Psi_L$ transforms as a $2$ irrep and $\widetilde{\Phi}_L$ as a $\bar{2}$ irrep (\cref{tab:quantum_numbers}), we expect to find $2 \otimes \bar{2} = 1 \oplus 3$ on the basis of the representation theory of \SU{2}; in other words the product of an \SU{2} doublet and conjugate \SU{2} doublet is the direct sum of the trivial and the adjoint representation. Before we proceed to demonstrate this well-known result, recall that the Pauli bivectors from \cref{eq:paulibivectors} can be written as
\[ \rrho_1 = \ebar{1}\ebar{2} \I, \quad \rrho_2 = \e{1}\ebar{2} \I, \quad \rrho_3 = -\e{2}\ebar{2} \I\ . \]
In order to verify $2 \otimes \bar{2} = 3 \oplus 1$ explicitly, we take
\[ \Psi_L = V \Box, \quad \Phi_L = W \Box, \]
where $V, W \in \GR[-]{3}$ are odd versors in \GR{3}. Explicit calculations yield
    \begin{align*}
        \Psi_L \widetilde{\Phi}_L = V \Box \widetilde{W} = V\widetilde{W} \, \I \ .
    \end{align*}
By the Cartan-Dieudonn\'e (CD) theorem (or equivalently, the closure of Clifford algebras), the product $V\widetilde{W} \in \GR[+]{3}$ is an even versor, i.e. a quaternion.
Hence, it is of the form
    \[ V\widetilde{W} = \alpha + \beta \e{1}\e{2} + \gamma \bl_1 + \delta \e{2}\ebar{1} \ , \]
and thus the product of this quaternion with the idempotent $\I$ equals
\begin{align*}
   V\widetilde{W}\I &= \bqty{\alpha + (\beta \ebar{1} - \gamma\e{2} + \delta\e{1})\ebar{2}}\:\I \ .
\end{align*}
This means that $V\widetilde{W}\I$ can be written in polar form as $\rho \exp(v\ebar{2})\I$ with $\rho \in \mathbb{R}$ a scale factor, which can be fixed to $\rho = 1$ after suitable normalisation of the spinors, and $v \in \mathbb{R}^3$. We thus find that the product of two left-handed leptons {\em almost} gives an element of the isospin \SU[I]{2}. As a matter of fact, the only thing that seems to be missing here is the `constant idempotent' $\Box = 1 - \I$ (see \cref{formula_exp_su2} for the Taylor expansion). This means that the behaviour of outor product states (like $\Psi_L \widetilde{\Phi}_L$) depends crucially on the chirality of the spinors it acts on. 
\begin{itemize}
    \item On even spinors in $S^+$, outor product states act trivially. Note that this is crucially different from saying that outor product states act as \SU[I]{2}, because the singlet states are trivially mapped to zero (annihilated). 
    \item On odd spinors in $S^-$, we know that $1 - \I = \Box$ has a trival action, which thus implies that 
    \[ \Psi_L \widetilde{\Phi}_LS^- = \big(\Box + \exp(v\ebar{2})\I\big)S^- = \SU[I]{2}S^-\ . \]
\end{itemize}
This can be summarised as follows\footnote{The physics of this statement might be a bit confusing at first, because it seems to predict a three fermion vertex, which are not allowed in the Standard Model.
However, we are currently only exploring the gauge sector, while ignoring the spacetime aspect, where the product of two fermions will be a boson after all.
}:
\[ \Psi_L \widetilde{\Phi}_L\big(\Theta_L + \Theta_R\big) = \Psi_L \widetilde{\Phi}_L\Theta_L = \Psi_L \widetilde{\Phi}_L\I\big(\Theta_L + \Theta_R\big)\ . \]
Equivalently, hereby taking a suitable normalisation into account, one can say that $\Psi_L \widetilde{\Phi}_L = V\widetilde{W}\,\I \in \SU[I]{2}\I$, where the idempotent at the right functions as a projection operator $S \rightarrow S^-$ on the odd spinor space. This type of argument can be repeated for all combinations.

Connecting back to the particle physics, it can straightforwardly be shown that the product of two left-handed neutrinos (of the form $\nu_L = \e{2} \Box z$) or two left-handed electrons (of the form $\varepsilon_L = \e{1} \Box z$) can only couple to $\I$ and $\rrho_3 = \bl_1 \I$, both of which commute with $Q = \bl_1$ and thus form a neutral current, as they must by charge conservation.
Similarly, the product of a left-handed electron and left-handed neutrino only couples to $\rrho_1$ and $\rrho_2$, which do not commute with $Q$ and hence form a charged current, as they must.

\subsubsection*{Products between isospin singlets}
Next we look at the outor product between two isospin singlets $\Psi_R, \Phi_R \in S^{+}$.
Given that $\Psi_R = \psi \Box$ and $\Phi_R = \varphi \Box$ with $\psi, \varphi \in \GR[+]{3}$ we find  
    \[ \Psi_R \widetilde{\Phi}_R = \psi \Box \widetilde{\varphi} = \psi \widetilde{\varphi} \Box \in S^+ \ . \] 
Following similar arguments to those given above for the outor product between isospin doublets, we find
\[ \Psi_R \widetilde{\Phi}_R\big(\Theta_L + \Theta_R\big) = \Psi_R \widetilde{\Phi}_R\Theta_R = \Psi_R \widetilde{\Phi}_R\Box\big(\Theta_L + \Theta_R\big)\ , \]
from which we conclude that $\Psi_R \widetilde{\Phi}_R \in \SU[Y]{2}\Box$ (after a suitable normalisation), where $\Box$ appears as the projection operator $S \rightarrow S^+$. 

To the best of our knowledge the appearance of an \SU[Y]{2} group from the (outor) product of two right-handed leptons is new,
but its appearance could make perfect sense since the isospin boson $W_\mu^3$ and hypercharge boson $B_\mu$ are not the physical degrees of freedom of the theory: the neutral gauge boson $Z_\mu$ and the photon $A_\mu$ are. These physical particles are obtained by rotating $W_\mu^3$ and $B_\mu$ over the Weinberg angle $\theta_W$ into $Z_\mu$ and  $A_\mu$, which in turn means that the corresponding isospin generator $\rrho_3$ and hyperspin generator $\rrho_0$ are mixed to form $I_3 - Q \sin^2 \theta_W$ and $Q$ respectively \cite{Nagashima:ch1}.
However, $\rrho_3$ comes from $\Psi_L \widetilde{\Phi}_L$ while $\rrho_0$ comes from $\Psi_R \widetilde{\Phi}_R$, and so there might be a (yet to be discovered) relationship between hyperspin and the Weinberg angle.

\subsubsection*{Products between isospin singlets and doublets}

Next we consider the outor product between an isospin doublet $\Psi_L = v q \Box \in S^{-}$ and an isospin singlet $\Phi_R = p \Box \in S^+$, with $p, q \in \GR[+]{3}$ bireflections and $v$ in $\mathbb{R}^3$ an extra reflection.
Then we find
    \[ \Psi_L \widetilde{\Phi}_R = v q \Box \widetilde{p} = v q \widetilde{p} \Box \in S^- \ . \]
Using the CD theorem, this product of 5 reflections can be reduced to a trireflection $v q \widetilde{p}$ which means that $\Psi_L \widetilde{\Phi}_R$ is an element of $S^-$. We then have that
\[ \Psi_L \widetilde{\Phi}_R\big(\Theta_L + \Theta_R\big) = \Psi_L \widetilde{\Phi}_R\Theta_R = \Psi_L \widetilde{\Phi}_R\Box\big(\Theta_L + \Theta_R\big)\ . \]
Once again using \cref{formula_exp_su2}, we get $\Psi_L \widetilde{\Phi}_R \in \GR[(1)]{3}\SU[Y]{2}\Box$. 
Note that apart from the projector $\Box : S \rightarrow S^+$ and the hyperspin group, an extra reflection in $\mathbb{R}^3$ appears here.
Since the outor product state $\Psi_L \widetilde{\Phi}_R$ is an element of $S^-$, we recognize that it must be an anti-Higgs boson $\widetilde{\phi} = \widetilde{\phi}^0 + \widetilde{\phi}^+$.

Finally, we must also consider the product between an isospin singlet $\Psi_R = q \Box \in S^{+}$ and an isospin doublet $\Phi_L = p v \Box \in S^-$.
In this case we find
    \[ \Psi_R \widetilde{\Phi}_L = q \Box v \widetilde{p} = \Box q v \widetilde{p} = q v \widetilde{p} \I \ . \]
This product of 5 reflections can again be reduced to a trireflection (times the idempotent $\I$), which means that
\[ \Psi_R \widetilde{\Phi}_L\big(\Theta_L + \Theta_R\big) = \Psi_R \widetilde{\Phi}_L\Theta_L = \Psi_R \widetilde{\Phi}_L\I\big(\Theta_L + \Theta_R\big) \]
Invoking \cref{formula_exp_su2}, we conclude that $\Psi_R \widetilde{\Phi}_L \in \GR[(1)]{3}\SU[I]{2}\I$, where not only the projector $\I : S \rightarrow S^-$ appears but also an additional reflection. This element corresponds to a Higgs boson $\phi = \phi^0 + \phi^+$.

\subsection{All degrees of freedom are accounted for}
A simple counting of degrees of freedom (DOF) shows that all $16$ DOF of \GR{4} have now been accounted for.
Starting with $S$ itself, which has $8$ real DOF. 
Next, $\widetilde{S}$ only contributes $4$ real DOF, since $\widetilde{S}^{(0)} \cong S^{(0)}$ and $\widetilde{S}^{(2)} \cong S^{(2)}$ and so these are not independent DOF, but $\widetilde{S}^{(1)}$ contains $4$ new real DOF.
Finally,  $\mathbb{R}\times\SU[I]{2}$ contributes the remaining $4$ real DOF ($3$ for the Lie group, and $1$ for the scalar factor $\rho$ from the previous section).
Hence, we have identified all $8 + 4 + 4 = 16$ DOF of the real GA \GR{4}. They can be summarised as follows: 
    \begin{table}[!h]
        \centering
        \begin{tabular}{l|c|c|}
            \cline{2-3}
            & $\Box$  & $\I$\\
            \hhline{~==}
            \rule{0pt}{2.5ex} $\GR[+]{4}$ & \multicolumn{1}{c|}{\cellcolor{nuRcolor}$S^+ \cong \widetilde{S}^+$} & \multicolumn{1}{c|}{\cellcolor{Xcolor}$\mathbb{R} \times \SU[I]{2}$}\\
            \cline{2-3}
            \rule{0pt}{2.5ex} $\GR[-]{4}$ & \multicolumn{1}{c|}{\cellcolor{nuLcolor}$S^-$} & \multicolumn{1}{c|}{\cellcolor{headercolor}$\widetilde{S}^-$} \\
            \cline{2-3}
        \end{tabular}
    \end{table}
This table should be read as follows: you can choose an element of $\GR[\pm]{4}$ and multiply this with either $\Box$ or $\I$. 
Viewed through the lens of \SU[I]{2}, the real algebra \GR{4} looks very different compared to when viewed through the lens of \Spin{4}.

\section{Particles of the Electroweak Sector}\label{sec:smfermions}

As we have seen in the previous sections, focussing on the gauge sector and excluding the spacetime sector, the elements of $S$ have exactly the quantum numbers we expect from the Standard Model leptons of the electroweak sector, while the generators of \un[I]{2} behave exactly like the hypercharge and isospin bosons. Moreover, the interactions between them follow the observed interactions in the electroweak sector.
To recapitulate, we identify the Standard Model particles as follows:
    \begin{alignat*}{2}
        \nu_R &= \Box z, \quad &\nu_L &= \e{2} \Box z, \\
        \varepsilon_L &= \e{1} \Box z, \quad &\varepsilon_R &= \e{12} \Box z, \\
        B_\mu &= B^0_\mu \rrho_0, \quad &W_\mu &= B^i_\mu \rrho_i
    \end{alignat*}
where all the $z = x + y \bl_1$ with $x, y \in \mathbb{R}$ for the leptons are `complex' numbers.
Here $\nu_{L/R}$ are left- and right-handed neutrinos, and $\varepsilon_{L/R}$ are left- and right-handed electrons, $B_\mu \in \un{1}$ is the gauge boson associated with hypercharge and $W_\mu \in \su{2}$ are the gauge bosons associated with isospin.
As we have shown in this paper, the leptons \emph{intrinsically} couple to hypercharge and isospin as expected: $\nu_R$ does not couple to hypercharge nor does it couple to isospin \SU{2}, and so describes the right-handed neutrino; $\nu_L$ has $I_3 = \tfrac{1}{2}$ and has hypercharge $Y = -1$, and so describes the left-handed neutrino; $\varepsilon_L$  has $I_3 = -\tfrac{1}{2}$ and has hypercharge $Y = -1$, and so describes the left-handed electron; and $\varepsilon_R$ is an isospin singlet with hypercharge $Y = -2$, and so describes the right-handed electron (see \cref{tab:quantum_numbers}).
Notice that the intrinsic coupling to hypercharge is in stark contrast to the usual approach, where the hypercharge eigenvalues are deduced from the known electric charge of the particles using the Nishijima–Gell–Mann formula $Q = I_3 + \tfrac{1}{2}Y$ \cite{Nagashima:ch1}, but do not follow from first principles.
A general lepton $\ell \in S$ can now be understood as the linear combination
    \begin{equation}
        \ell = \nu_L + \nu_R + \varepsilon_L + \varepsilon_R \ ,
    \end{equation}
with squared norm 
    \begin{equation}
        \widetilde{\ell} \ell = \widetilde{\nu}_L \nu_L + \widetilde{\nu}_R \nu_R + \widetilde{\varepsilon}_L \varepsilon_L + \widetilde{\varepsilon}_R \varepsilon_R \ .
    \end{equation}
In order to decompose into left- and right-handed spinors, we can simply use grade projection onto the even and odd parts of \GR{4}:
    \begin{equation}
        \ell_R \coloneqq \expval{\ell}_+ = \nu_R + \varepsilon_R \in S^{+}, \quad \ell_L \coloneqq \expval{\ell}_- = \nu_L + \varepsilon_L \in S^- \ .
    \end{equation}
Equivalently this decomposition could be done using the chiral projectors $\tfrac{1}{2} (1 \mp \Gamma)$ where $\Gamma = \bl_1 \bl_2$ since the odd elements anti-commute with $\Gamma$ whereas the even commute with $\Gamma$.
Alternatively we could decompose $\ell$ into electron and neutron states as
    \begin{equation}
        \nu = \tfrac{1}{2} (\ell + Q \ell \widetilde{Q}), \qquad \varepsilon = \tfrac{1}{2} (\ell - Q \ell \widetilde{Q})\ .
    \end{equation}
A final decomposition that may be of interest is how, given that any lepton can be written as $\ell = X \Box$ with $X \in \GR{3}$, $X$ can be recovered since $\Box$ is non-invertible. The solution is to realize that all the elements of \GR{3} live in a subspace defined by $\e{12}\ebar{1}$, and so all elements containing $\e{2}$ will anti-commute with $\e{12}\ebar{1}$ and hence \emph{an} $X$ satisfying $\ell = X \Box$ can be recovered as
    \begin{equation}
        X = \ell + \e{12}\ebar{1} \ell \ebar{1}\e{21} \ .
    \end{equation}
The existence of all these decompositions should not distract us from the fact that the main actor in $S$ is the lepton $\ell$, and that $\ell_1 \widetilde{\ell}_2$ is the (unnormalized) transformation from $\ell_2$ to $\ell_1$, regardless of how these breakdown into irreps.
This mindset has the potential to greatly simplify computations, which is something we will revisit in future work.

\section{Discussion}\label{sec:discussion}

In this section we will enumerate the main contributions of this work, and discuss their impact in a broader context.

\paragraph{Keep it real}{The approach taken in this paper is significantly different from the traditional approach taken by other authors, because it does not require complexification to define the spinor space.
As a reminder, the traditional approach to constructing a spinor space would involve the introduction of an imaginary unit, going from $\GR{4} \to \GC{4}$, in order to form a Witt basis $v_{\pm j} = \tfrac{1}{2}(\e{j} \pm i \ebar{j})$, also referred to as ladder operators, with which one can construct an idempotent 
    \[ \boxplus = \prod\nolimits_{j=1}^4 v_{+j}v_{-j} \ , \]
which is then used to define the complex spinor space $S_\mathbb{C} = \GC{4} \boxplus$, see text books such as \cite{Polchinski:1998rr,alma9912723332602411}, or N. Furey's work for a specific application to the electroweak case \cite{doi:10.1142/S0217751X18300053}.
In contrast, we formed a real spinor space simply by asking a very geometric question: what would happen if for whatever reason we could not distinguish the different commuting rotations in our space? This led to the forgetful idempotent
    \[ \Box = \tfrac{1}{2} (1 + \bl_1 \bl_2), \]
and a \emph{real} spinor space $S = \GR{4} \Box$.
In the language of the current work, the traditional $\boxplus$ is also a forgetful idempotent, satisfying the following defining forgetfulness relation:
    \[ i \boxplus = \bl_1 \boxplus  = \bl_2 \boxplus = -i\bl_1\bl_2 \boxplus \ . \]
As a result of this forgetfulness relation $\text{dim}(S_\mathbb{C}) = \text{dim}(S) = 8$ real DOF and so, somewhat counter-intuitively, the two spinor spaces have the same dimensionality despite complexification.
Put differently, adding $i$ does not convey any new information, because the same element is already represented by $\boxplus \bl_j $.
The one advantage that $i$ has however, is that as a scalar, it commutes with all the elements of \GR{4}, whereas $\bl_j$ does not share this property. This is why we had to be careful to write the `complex number' $z = x + y \bl_1$ on the right of $\Box$ in order to retrieve correct eigenvalues.

\paragraph{Spinors Transform two-sidedly}{
An important consequence of our analysis is that spinors $\Psi$ transform under spin transformations $U \in \Spin{4}$ just like everything else: under conjugation $\Psi \to U \Psi \widetilde{U}$. However, if the particular spin transformation happens to be an element $U \in \SU[I]{2} \subset \Spin{4}$ then spinors ignore $\widetilde{U}$ coming from the right since $\SU[I]{2}$ is the invariance group of $\Box$ and hence spinors appear to transform one-sidedly as $\Psi \to U \Psi$. But we saw that it was important not to take this as the definition of spinor transformations, in order to get the correct \UN{1} transformations.
The importance of breaking from two-sided to one-sided was also recently commented on by N. Furey \cite{Furey:2023wpi}, who called the breaking from $\Spin{2n} \to \SU{n}$ the ``multivector condition''.
We observe that our geometric approach explains the origin of this constraint, and makes the construction of minimal left ideals that obey the multivector condition straightforward. 
This will be the subject of an upcoming paper.
}

\paragraph{Spinor Spaces of \SU{n}}{
The seminal paper ``Lie Groups as Spin Groups'' derived how all the classical Lie groups can be formulated as subgroups of Spin groups \cite{LGasSG}. 
But their discussion of spinor spaces was restricted to spinor spaces of the general linear group within \GR{m,m}. Meanwhile, in their approach (which we followed) \SU{n} groups are subgroups of \Spin{2n,0} or \Spin{0,2n}, but their paper does not provide an answer on how the spinor space is to be constructed in all-positive or all-negative signatures.
The current work provides a glimpse of how this can be done, and in an upcoming paper we will provide the general argument for the construction of spinor spaces in \SU{p, q}.
The specific treatment of \SU{3} is also of great interest, given its role in the strong interactions.
A manuscript on this specific gauge group is underway, 
which will finally deliver on promises made in \cite{roelfs_su3},
and connects to work done on \SU{3} and octonions \cite{lasenby2022recentresultssu3octonions,furey2016standardmodelphysicsalgebra}.
}

\paragraph{The Desert of the Real}{
We have seen that while $\I$ corresponds to the identity matrix, $\Box$ corresponds to the zero matrix, and hence our formulation of the spinor space $S$ has no matrix equivalent. Instead, in the (matrix) representation theory of \SU{2} the spinors are represented as (complex) column vectors acted upon by \SU{2} matrices, which separates transformations from the elements being transformed). 
Contrarily, within GA transformations and elements live in the same space and are treated on equal footing.
As a result, we find that within the GA \GR{4}, the group \Spin{4} acting on \GR{4} itself naturally leads to the discovery of spinors as elements of a minimal left ideal, and the elements of this minimal left ideal happen to behave exactly like the irreps we need to represent chiral leptons.
This is a real victory for the geometric algebra approach to Lie Groups pioneered in \cite{LGasSG}, and it makes one wonder what this approach could bring to GUT theories, see \cite{baez2010algebra} for an excellent introduction to GUT theories.
}

\paragraph{Handedness is gradedness}{The difference between left- and right-chiral leptons is due to the graded structure of the algebra, with right-chiral corresponding to even and left-chiral corresponding to odd. 
This correspondence to grade resolves the question of why right-handed leptons do not interact with weak \SU[I]{2} or do not have a separate \SU{2} group of its own: geometry does not work that way.
This serves as a good example of the benefits the geometric mindset brings.
}

\paragraph{One Rep Max}{Irreducible representations play a crucial role in Gauge Theories, because there is overwhelming evidence that particles correspond to the irreps of certain symmetry groups. However, it is important not to lose track of the goal of gauge theory: to accurately describe particle physics. And when actually performing computations, describing everything in terms of irreps means the poor physicist has to do all the bookkeeping. What this paper aims to stress is that when computing you should go for \emph{One Rep Max}: use an algebra which naturally encodes all the physics for you and let the algebra take care of the bookkeeping.}

\section{Conclusions \& Outlook}

In conclusion, this novel approach to the construction of spinor spaces is able to reproduce the properties of spinor spaces without requiring complex numbers.
The current work hardly does this innovative approach justice. 
Several other manuscripts on the forgetful idempotent approach to spinor spaces are currently under preparation, such as the generalization to spinor spaces of arbitrary dimension and signature, the Cayley-Dickson construction, and the specific treatment of \SU{3}.

The current work only focused on the gauge sector. 
To include the spacetime dependence and obtain a full Yang-Mills theory, the spinor space derived in this paper needs to be promoted to a quantum field.
But before we can do that, we must first reevaluate Dirac spinors in spacetime through the lens presented by the ideas in this paper.
This will be the subject of the second paper of the lepton triptych.

\section*{Acknowledgements}
The authors would like to thank Steven De Keninck and Moab Croft for invaluable discussions about this research.

\bibliographystyle{abbrvunstrnat}
\bibliography{biblio.bib}

\end{document}